\documentclass[reprint,groupedaddress,superscriptaddress,floatfix,longbibliography,tightenlines,nobibnotes,nofootinbib,aps]{revtex4-2}

\usepackage[utf8x]{inputenc}
\usepackage[english]{babel}
\usepackage[T1]{fontenc}
\usepackage{lmodern}
\usepackage{dsfont}
\usepackage{mathtools}
\usepackage[margin=0.6in]{geometry}
\usepackage{fancyhdr}
\usepackage{amsmath,amsfonts,amssymb,amsthm,bm,times}
\usepackage{graphicx,color}
\usepackage[caption=false]{subfig}
\usepackage{textcomp}
\usepackage{lipsum}
\usepackage{verbatim}
\usepackage{blindtext}
\usepackage{natbib}
\usepackage{makecell}
\usepackage{ulem}
\usepackage{booktabs}
\usepackage[toc,page]{appendix}

\usepackage{microtype}

\usepackage{braket}

\begin{document}
\title{\textbf{Carbon nanotube as quantum point contact valley-filter and valley-splitter}}

\author{N. A. Hadadi}
\affiliation{National university of Singapore, Singapore}
\author{A. Belayadi}
	\affiliation{Ecole Supérieure des Sciences des Aliments et des Industries Alimentaires, ESSAIA, El Harrach, 16200 Algiers, Algeria}
	\affiliation{University of Science and Technology Houari Boumediene, Bab Ezzouar, 16111 Algiers, Algeria}
 \author{Ousmane Ly}
 \affiliation{Universite de Nouakchott, Facult ´ e des Sciences et Techniques, ´
Departement de Physique, Avenue du Roi Faic¸al, 2373, Nouakchott, Mauritanie}
	\author{A. Abbout}
	\email{adel.abbout@kfupm.edu.sa}
	\affiliation{Department of Physics, King Fahd University of Petroleum and Minerals, 31261 Dhahran, Saudi Arabia}
 \affiliation{Quantum Computing Group, Intelligent Secure System (ISS) center,  King Fahd University of Petroleum and Minerals, 31261 Dhahran, Saudi Arabia}



  \begin{abstract}
  The electrical characteristics of a carbon nanotube can be significantly modified by applying elastic strain. This study focuses on exploring this phenomenon in a single-walled carbon nanotube (SWNT) using tight-binding transport calculations. The results indicate that, under specific strains, an armchair SWNT can act as a filter, separating the two valley electrons K and K$'$. Notably, when subjected to deformation, the SWNT exhibits intriguing behaviors, including a quantized conductance profile that varies with the strength of the strain. Consequently, precise control of the width of the quantized plateaus allows for the generation of a polarized valley current. Furthermore, when both K-types are conducted, the strain is demonstrated to completely separate them, directing each K-type through a distinct pathway.
  \end{abstract}
\maketitle

\section{INTRODUCTION}
The potential of graphene for application in carbon electronics lies in the possibility of fabricating devices with the two valley electron types K and K$'$ \cite{neto2009electronic,novoselov2005two}, which have no analog in silicon-based electronics. A relatively new field of nanoelectronic application known as valleytronics has emerged as a result of graphene's spin-like valley degree of freedom which aims to design high-efficiency and low-dissipation electronic devices. One of the main requirements of valleytronic applications is to have a simple and effective method for generating and controlling valley-polarized currents. In other words,  a controlled way of manipulating a single valley type in graphene and the generation of valley-polarized currents are required. Similar to the uses of the electron spin in spintronics or quantum computing, the independence and the degeneracy of the valley degree of freedom indicate that it could be employed for this purpose.

Due to its distinct physical and electrical characteristics, such as its great strength and stiffness, Carbon Nanotube (CNT) has been widely regarded as a very promising material for uses in nano-engineering ever since its discovery \cite{iijima1991helical,yengejeh2016advances}. It is widely acknowledged that a CNT's properties change greatly when it is mechanically deformed. In addition to this coupling, CNTs are known to have remarkable flexibility when subjected to external hydrostatic pressure and bending force which makes it a very  interesting material to investigate. However, they normally break at strains of less than 15\% at room temperature, and in practice, only 6\% is attained because of structural flaws \cite{sato2011elastic, zhang1998plastic}. This means that if the curvature strength in a CNT goes up to less than the level of the structural failure, the behavior of the material can be predicted to be elastoplastic. However, in this paper, it is assumed that the CNT can sustain its elastoplasticity for strong bending, hence the breaking problem was not considered.  Moreover, the electromechanical coupling or the interrelationship of the electronic and mechanical properties of single-walled CNT (SWNT) was demonstrated \cite{rochefort1998effect,bozovic2001electronic} and it was found that the reversible bending of a CNT  can be used to alter its conductance where as the bending increases the conduction through the nanotube reduces \cite{maiti2002electronic,tombler2000reversible}. Previous studies as well showed that the conductance of metallic SWNTs (armchair CNT),  which is the main type we investigate in this work,  should be insensitive to minor bending deformations up to a certain angle, and reduction should appear after surpassing this bending angle \cite{rochefort1999electrical,kane1997size}. 

In accordance with their chirality, SWNT can be metallic with no bandgap (Armchair), semiconducting with a significant bandgap (Zigzag), or in between those two cases (Chiral)  which can determine  their scattering mechanisms \cite{ilani2010electron, alfonsi2008small}.  This paper will show that the Armchair SWNTs can display a quantized electronic conductance profile under mechanical bending and can be exploited to be a valley filtering device in a very controllable way by suitable mechanical deformations. In addition, it will be demonstrated that the bent Armchair SWNT interestingly manifests a clear splitting of the two valley types by forcing them to take two unique trajectories through the strain region. These phenomena are  not manifested in the other two SCNT types. 



\section{Strained Carbon Nanotube}
\begin{figure}[h!]
\centering
  \includegraphics[width=1\linewidth]{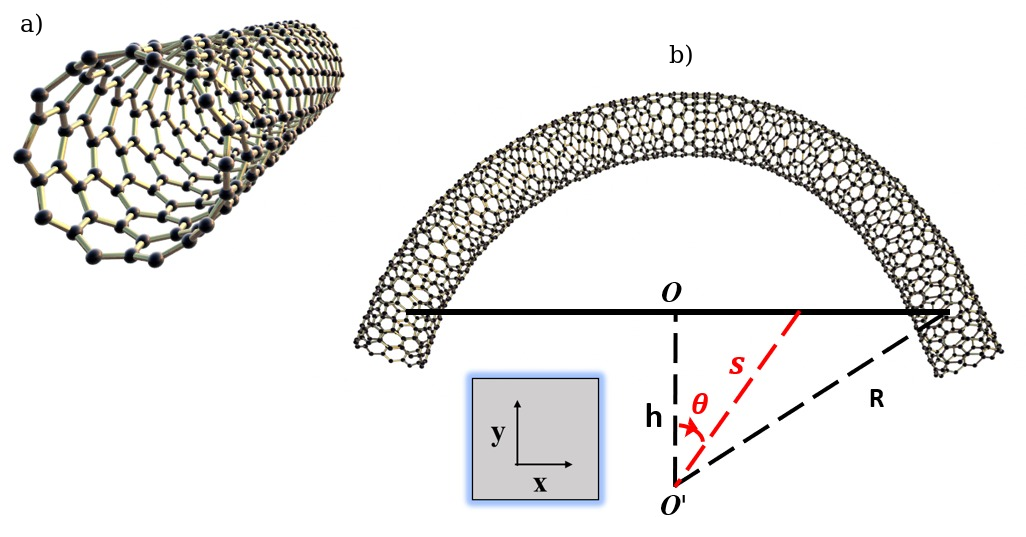}
  \caption[]{a) A straight carbon nanotube system. b) A curved CNT with a center of curvature $O'$ that is $h$ units down the previous center $O$. Each cross section of the CNT with a center identified by an angle $\theta$ will be rotated to be along $\overrightarrow{O'S}$ and then translated to the circular deformation positions (On a circle).  }
  \label{fig:nanotube}
\end{figure}


The SWNT system was initially created by a super unit cell with the unit vectors \textbf{T} and \textbf{U}:
\begin{equation}
  \textbf{U} = n \textbf{u}_1 + m \textbf{u}_2
\end{equation}
\begin{equation}
 \textbf{T} = \frac{-(2m+n)\textbf{u}_1+(2n+m)\textbf{u}_2}{gcd(2m+n,2n+m)}     
\end{equation}

Where \textbf{U} and \textbf{T} are the chiral vector and transitional vector, respectively. \textbf{u}$_1$ and \textbf{u}$_2$ are  the lattice basis vectors of graphene, (n, m) represents the chiral indices and they are positive integers, and gcd is the greatest common divisor function. The vector \textbf{U} gives the circumference of the CNT after rolling it up. Thereby, the SWNT's radius  is given by $r =\sqrt{3}\sqrt{n^2+nm+m^2}/(2\pi)$.

When subjected to an external load that exceeds a critical threshold, carbon nanotubes experience buckling, resulting in a transformation of their shapes and consequently, their mechanical characteristics. Additionally, factors such as differential thermal expansion or the presence of imperfections can trigger a similar response. This deformation induces alterations in both the mechanical and physical attributes of CNTs. Furthermore, this deformation causes localized variations in the spacing between carbon atoms, leading to a strain that can generate an intense pseudo-magnetic field akin to what is observed in graphene nano-bubbles. What makes carbon nanotubes particularly intriguing is that this deformation at the point of maximum buckling exhibits contrasting behaviors between their upper and lower surfaces. Specifically, the curvature of the deformation along the two principal symmetry axes is positive for the upper surface but opposite in sign for the lower surface. Consequently, the quantum transport of electrons within CNTs is profoundly impacted and warrants thorough investigation.\\

Deviation from a straight linear form of carbon nanotubes to a curved configuration can be achieved through various experimental methods \cite{ShimaMaterial}. The axial buckling method involves applying axial compressive forces, resulting in displacement proportional to the force within the elastic region \cite{Kuzumaki_2006,Misra_2007}. Several approaches, including continuum modeling and molecular dynamics simulations, exist to characterize the shape and behavior of CNTs \cite{ARASH2012303}. The buckling method induces diverse final shapes with additional possible imperfetions such as radial corrugations or surface rippling. In our manuscript, we assume that our deformed CNT maintains a circular cross-sectional area without rippling or surface corrugation, and we limit the effect of strain on the CNT in the alteration of hopping parameters between carbon sites. Consistent with references \cite{settnes2016graphene,Naif2023} on nanobubbles and \cite{Andrei} on twisted bilayer graphene, we disregard any alteration in the electronic atomic structure of carbon atoms.

In this manuscript, we explore a deformation involving a segment of length $L$ within a carbon nanotube, transforming it into a circular shape with a radius $R$, as illustrated in Figure \ref{fig:nanotube}. The remaining section of the CNT serves as leads and remains linear, aligned with the final rotated cross-sections of the system. The newly formed CNT's center, denoted as $O'$, is positioned at a distance $h=\sqrt{R^2-(\frac{L}{2})^2}$ relative to the original reference point, $O$. We implement this CNT transformation following a specific model. Any given site on a transverse cross-section of the CNT undergoes an initial clockwise rotation by an angle $\theta$ about the local z-axis (the axis passing through the point $S$), followed by a translation of length $(R-s)$ in the direction of vector $\overrightarrow{O^\prime S}$. The angle $\theta$ depends on the position $x$ and is calculated as $\theta=\arctan{\frac{x}{h}}$, while the distance $s$ is determined as $s=\frac{h}{\cos(\theta)}$. The new position $(x',y',z')$ of a carbon atom becomes:

\begin{equation}
\begin{dcases}
x'= &x+(y+R-s)\sin\theta \\
y'= &  (y+R-s)\cos\theta\\
z'=z
\end{dcases}
\label{eq:R}
\end{equation}
 As a result of this deformation, the hopping parameter in the Hamiltonian will be modified by
\begin{equation}
   t_{ij} = t_0e^{-\beta (\frac{d_{ij}}{d_0}-1)}
   \label{eq:hopping}
\end{equation}
 Where $d_{ij}$ represents the length of the strained lattice vector, $t_0 = 2.7 \text{ eV}$, $\beta = 3.37$, and $d_0 = 0.142$ nm is the length of the unstrained C – C bond for graphene structure \cite{stegmann2018current,Naif2023,kleinherber}.

In experimental settings, the buckling of a carbon nanotube can be achieved through the application of mechanical force perpendicular to its principal axis \cite{Misra_2007,ShimaMaterial} , by inducing thermal expansion or contraction \cite{PAYANDEHPEYMAN2019258}, or by utilizing a substrate with a specific pattern or shape \cite{khang}. In these scenarios, it is anticipated that there will be variations in the electronic density of states. However, in our model, we choose to overlook this change and focus exclusively on the modification of the hopping parameter as the predominant factor influencing the system.

\section{SWNTs$'$ Conductance Profile}

In order to obtain a valley-polarized current one needs to address the two valleys in  the system individually as independent degrees of freedom of the conduction electrons. In our SWNT's system, we examine how electronic conductance behaves. We achieve this by injecting electrons from one lead, connected from one end, to another lead at the opposite end. We utilize a tight-binding model and employ Kwant \cite{Groth_2014}, a numerical tight-binding calculation package, to effectively analyze the transmission probability between these two leads under various bending conditions. Notably, the conductance profile can vary significantly based on the type of nanotube or its chiral indices (n, m). Three well-studied types include zigzag (n, 0), armchair (m=n), and chiral nanotubes. In this investigation, we focus on the armchair type, as it proves to be the most suitable for filtration purposes due to its capability to easily distinguish between the two valleys. We also account for the influence of strain by altering the curvature parameter $R$ in Eq. \ref{eq:R} , as depicted in Figure \ref{fig:nanotube}. The initial SWNT has zero curvature (infinite radius $R$). Decreasing the parameter $R$ leads to increased bending, while increasing it results in the opposite effect.

 \begin{figure}[b!]
\centering
  \includegraphics[width=0.9\linewidth]{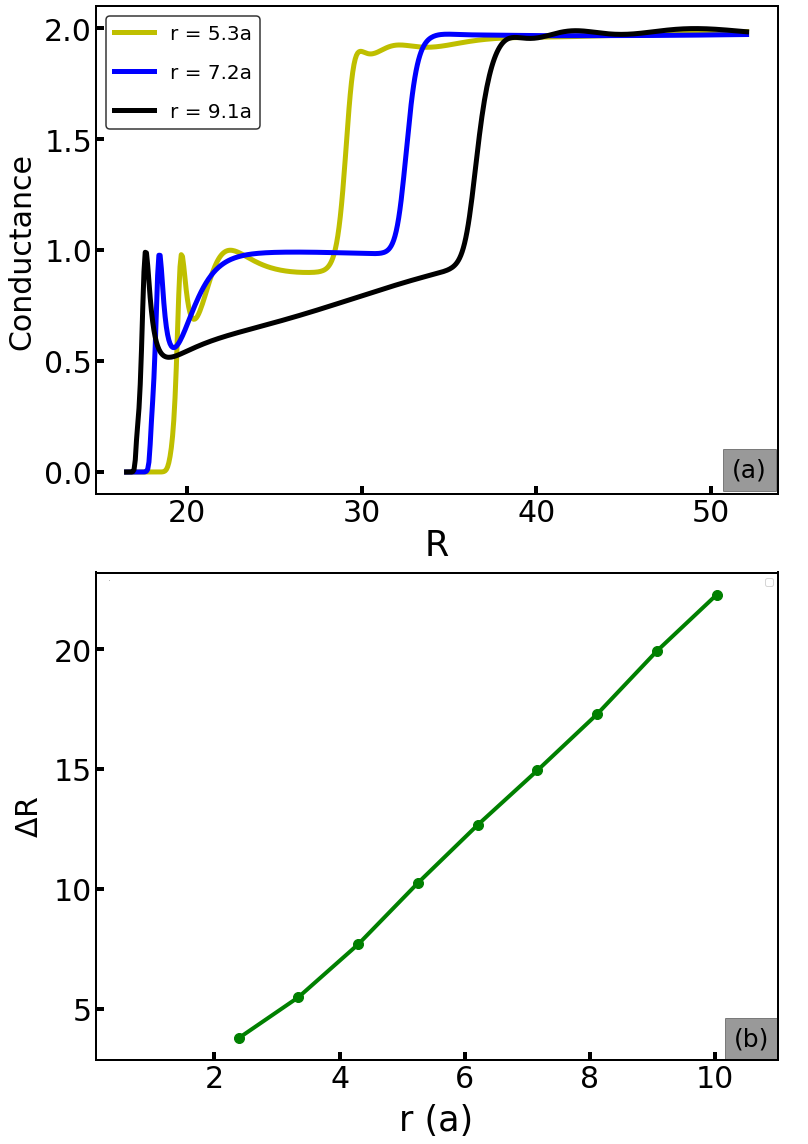}
  \caption[]{\textbf{(a)} Conductance versus the curvature parameter R with different values of nanotube radius r. The length of the nanotube was held fixed with a value L = 18.5a.  \textbf{(b)} $\Delta R$ the width of the first plateau versus the length of the nanotube. $\Delta R$ = $R_2 - R_1$ where $R_1$ is the beginning of the first plateau (first mode) in the conductance as a function of R (as shown in plot \textbf{(a)})and  $R_2$ is value of R at the end of the first plateau. \textbf{(c)} The insert plot shows how $R_1$ and $R_2$ change the opening $\Delta R$. }
  \label{fig:opening_r}
\end{figure}

The conductance profile of the SWNT as a function of the curvature parameter $R$, which characterizes the strength of strain is depicted in Fig. \ref{fig:opening_r}a. Here, we consider only the first two modes $K$ and $K'$, for which the Fermi  energy is set to $E_\text{F}= $ 0.15t$_0$. The nanotube's length is $L = 18.5a$ where  $a = d_0 \sqrt{3}$. In figure . \ref{fig:opening_r}a) the plots  illustrate the conductance with different nanotube radii r. Observations reveal that when the system remains unstrained  (for large R, e.g. when $R>40a$ and the nanotube radius r = 7.2a in Fig. \ref{fig:opening_r}a )   the current is unpolarized and both valleys  flow along the system and are fully transmitted. However, as the bending becomes more pronounced,  the $K$-type valley electrons are selectively suppressed, leading to the emergence of a perfectly polarized current (e.g. when $17a<R<30a$ and r = 7.2a).
Subsequently, as the bending intensifies further, the electrons from the other valley are also effectively filtered out (e.g., $R < 17a$ and r = 7.2a in Figure  \ref{fig:opening_r}a).\\
The filtering effect in the SWNT (selecting one valley) depends on the strength of the deformation. In order to capture this effect, we calculate the conductance for various SWNT radii and report the quantity $\Delta R=R_2-R_1$ where $R_1$ is the deformation radius at the opening of the first plateau of conduction and $R_2$ is the corresponding value at the opening of the second plateau of conduction as shown in figure \ref{fig:opening_r}a. According to this definition, $\Delta R$ represents the length of the interval over which the current is polarized by valley.We can clearly identify in figure \ref{fig:opening_r}b a linear relationship between $\Delta R$ and the size of nanotube and deduce that the filtering effect is better (occurring over larger deformation intervals) in the case of larger-diameter nanotubes.

\begin{figure}[b!]
\centering
  \includegraphics[width=1\linewidth]{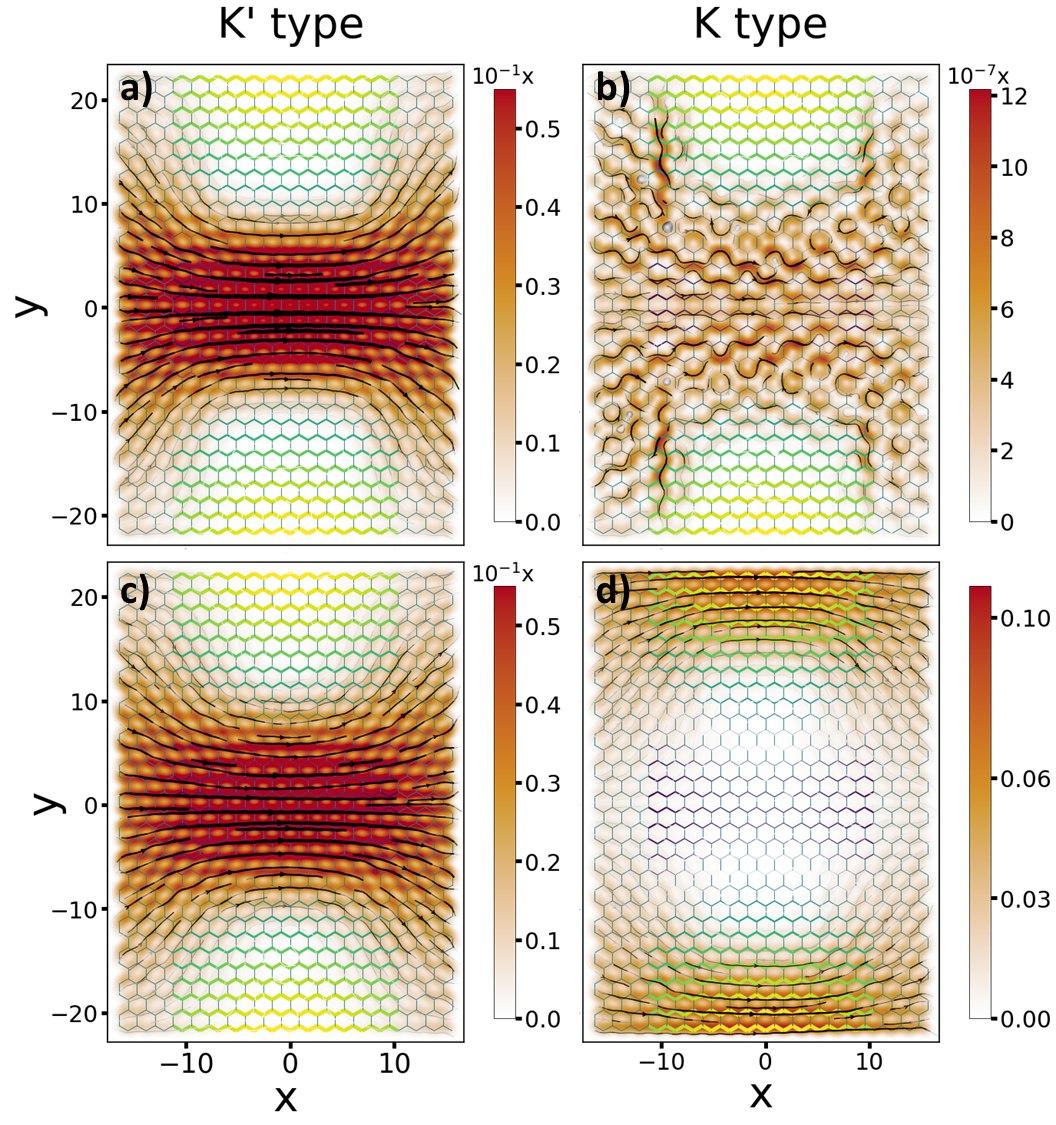}
  \caption[]{The current maps through the nanotube system with $E_\text{F} = 0.15t_0$ and only two modes $K$ and $K'$. All the maps have the same width and height. The nanotube's length and radius are 18.5a and 7.15a, respectively.   The columns represent the maps for each valley type. (\textbf{a}, \textbf{b}) plots illustrate the maps for the first plateau in the conductance versus R (R = 25a) plot (Fig. \ref{fig:opening_r}a), while (\textbf{c}, \textbf{d}) were plotted for $R = 45a$. IN figure textbf{b} the magnitude of the current is 6 order of magnitude  smaller than that of $K'$ thus we can  claim that we have a filtering effect. }
  \label{fig:currents}
\end{figure}

Fig. \ref{fig:currents} illustrates the current distribution of individual valley electrons, $K$ and $K'$ in 2D. To obtain these current maps, the SWNT is kept undeformed and the impact of strain is solely considered in the hopping parameters. Finally, we flatten (unroll) the SWNT to create a 2D map.



The parameters we used to obtain the  2D current maps (Fig. \ref{fig:currents}) are E$_\text{F}$ = 0.15t$_0$,  $L = 18.5a$, and $r = 7.15a$ (the same as for the blue plot in Fig. \ref{fig:opening_r}a ) for different bending strengths controlled by the curvature parameter R. For  R = 45a and r = 7.2a, the conductance is in the second conduction plateau as shown in  Fig. \ref{fig:opening_r}. Under these specific conditions, the current maps reveal that both  $K$ and $K'$ electrons are transmitted (Fig. \ref{fig:currents} (c-d)). Intriguingly, the $K$ and $K'$ electrons are perfectly separated while passing through the system. This splitting shows one of them passing through the middle while the other propagates through the two edges of the 2D map. This behavior highlights that nanotube functions akin to a quantum point contact (QPC), with the bending serving as a simple mechanism for opening and closing the constriction \cite{AdelAbbout,Wees}. Furthermore, the emergence of the quantized plateaus  serves as a distinctive characteristic of QPC behavior. This phenomenon can be comprehended by considering that the bending alters the distances between sites on the lateral surface differently than on the central surface of the region of buckling in the nanotube. Consequently, these alterations create constrictions that enable electron passage selectively according to their wavelengths.  It is worth noting that in addition to the splitting effect (figure \ref{fig:currents} (c) and (d)) , the nanotube serves as a filter for large bending (small R) by letting only one valley (figure \ref{fig:currents} (a)) while the second one is almost blocked (6 orders of magnitude smaller in figure \ref{fig:currents}(b))

Finally, when the strain gets too strong (e.g., when $R<16a$ in Fig. \ref{fig:opening_r}a and r = 7.2a), both types are entirely switched off. This clearly indicates that the strain strength can thoroughly dictate the conductance behavior and, consequently, can be employed as a control tool for the valleys' transmission. The colors in the current maps (specifically, the green-yellow in the edges and the dark blue in the middle) illustrate the strength of the modified hopping parameter t$_{ij}$ as in Eq. \ref{eq:hopping}. The green-yellow regions  in Fig. \ref{fig:currents} are showing the stretched part of the strained nanotube (the top of the nanotube) while the dark-blue region in the middle of Fig. \ref{fig:currents} shows where the hopping parameters are contracted (the bottom of the nanotube). 


One of the main reasons for the filtering and splitting of the valley's electrons seems to arise due to the interaction of the electrons with the nanotube deformation.  These phenomena could happen as a result of the induced Pseudo Magnetic Field (PMF) that is produced when hexagonal graphene structure is  strained \cite{settnes2016graphene,milovanovic2016strain}. This PMF acts oppositely on the two valleys where the generated PMF  causes the $K$ and $K'$ electrons to follow different real space trajectories. 
The plateaus in conductance profiles are not well pronounced (Fig. \ref{fig:R_L}a) at least at the beginning of the first plateau where there are visible oscillations. In fact, it can be seen (Fig. \ref{fig:R_L}a) that as the length L of the nanotube increases  the oscillations in the plateaus increases. This behavior is well understood as the interference of the wave functions back-scattered from the beginning and the end of the QPC region \cite{AdelAbbout2}.  Interestingly, the bent SWNT can be more advantageous than the QPC in two dimensional electronic gases  because when both valley types are conducted, they are well separated (Fig. \ref{fig:currents}c-d).   
 Thus, in addition to providing a filtering method, the strain SWNT also gives a way to separate them effectively. 
 
\section{SWNT's Parameters Effect on Valley Splitting and Filtering}
\begin{figure}[t!]
\centering
  \includegraphics[width=0.9\linewidth]{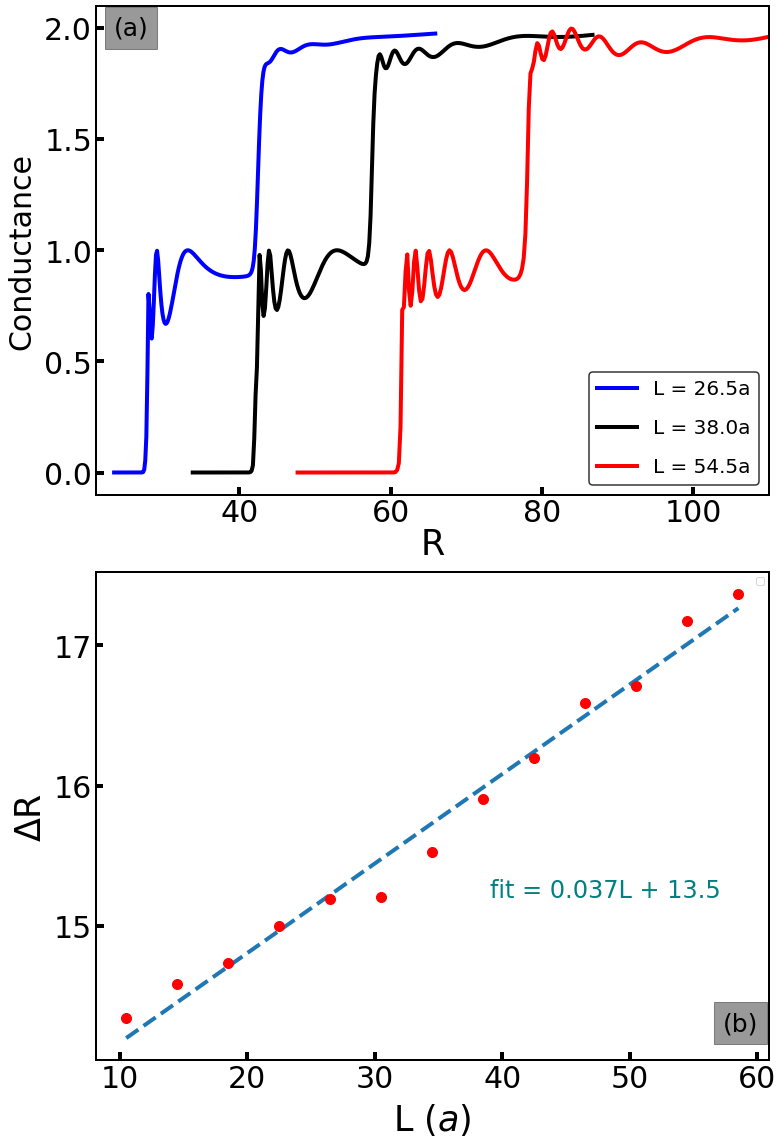}
  \caption[]{\textbf{(a)} Conductance versus the curvature parameter R with different values of nanotube length L. The radius of the nanotube was held fixed with the value r = 7.2a, and $E_F$ = 0.15 eV. \textbf{(b)} $\Delta R$ the width of the first plateau versus the length of the nanotube. $\Delta R$ = $R_2 - R_1$ where $R_1$ is the beginning of the first plateau (first mode) in the conductance as a function of R (plot \textbf{(a)}) and  $R_2$ is value of R at the end of the first plateau. }
  \label{fig:R_L}
\end{figure}
In our system, there are several parameters that can affect the conductance profile and, hence, the filtering process which are the nanotube radius r, the length of the nanotube L, and the opening $\Delta$R (which is the width of the first plateau in Fig. \ref{fig:R_L}a or Fig. \ref{fig:opening_r}a in terms of R). In fact, the opening parameter, which is the most important parameter since it determines the filtering and the spatially splitting process,  is affected by both the radius r and the nanotube length L. Therefore, it is reasonable to focus on the opening $\Delta$R to study the effects of the SWNT's parameters on the performance of the device. To investigate the effect of r on  $\Delta$R, the conductance versus the curvature parameter R is plotted with various values of r while fixing  L as shown in Fig. \ref{fig:opening_r}a. Fig. \ref{fig:opening_r}a shows that the opening increases as r increases. Additionally, the increase in the separation between the beginning of the two plateaus is perfectly linear as shown in Fig. \ref{fig:opening_r}b.  On the other side, Fig. \ref{fig:R_L}a demonstrates the conductance profiles while changing L and holding r as constant where r was chosen to be 7.2a which gives the optimal conductance profile as shown in Fig. \ref{fig:opening_r}. Similarly, the opening $\Delta$R also increases almost linearly as L increases. This manipulation of the opening length can pave the way to design a valley filtering and splitting  device with easily controllable valley polarization.




\section*{CONCLUSION}
In this paper, we showed that under particular strains, an armchair single-walled carbon nanotube (SWNT) can play the role of filtering and splitting the two valley electrons K and K$'$. It was demonstrated that the SWNT can manifest interesting phenomena under these conditions, where the electronic conductance was quantized when plotted as a function of the strain strength. There were two pronounced plateaus that correspond to the transmission of each K-type electron. Moreover, the conductance profile showed a large width of the first plateau ($\Delta $R) where the K$'$ type is passed and the K-type completely suppressed, and this width is significantly affected by the SWNT's parameters such as the length, and the radius of the nanotube. Finally, the relations of these parameters with the opening $\Delta$R are perfectly linear in terms of the radius parameter (Fig.\ref{fig:opening_r}b) and almost linear in the case of the length of the nanotube (Fig.\ref{fig:R_L}b). This ability to filter and split the K-types without the need for outside intervention after the strain could foster the development of valley transistors.

\section*{ACKNOWLEDGMENT}
The authors acknowledge computing time on the supercomputer SHAHEEN at KAUST Supercomputing
Centre and the team assistance. A.A. gratefully
acknowledge the support provided by the Deanship of
Research Oversight and Coordination (DROC) at King
Fahd University of Petroleum and Minerals (KFUPM)
for funding this work through the Intelligent secure system center research grant No. INSS2401.


\bibliographystyle{IEEEtran}
\bibliography{reference}


\end{document}